\documentclass[%
 aip,
 jmp,%
 amsmath,amssymb,
 reprint,%
]{revtex4-1}

\usepackage{graphicx}
\usepackage{dcolumn}
\usepackage{bm}
\usepackage{multirow}
\usepackage{upgreek}

\begin{document}

\preprint{AIP/123-QED}

\title{High-resolution, vacuum-ultraviolet absorption spectrum of boron trifluoride}

\author{Patrick P. Hughes}
\affiliation{\mbox{National Institute of Standards and Technology, Gaithersburg, Maryland, 20899, USA}}
\author{Amy Beasten}
\affiliation{\mbox{Nuclear Engineering Program, University of Maryland, College Park, Maryland, 20742, USA}}
\author{Jacob C. McComb}
\affiliation{\mbox{Nuclear Engineering Program, University of Maryland, College Park, Maryland, 20742, USA}}
\author{Michael A. Coplan}
\affiliation{\mbox{Institute for Physical Science and Technology, University of Maryland, College Park, Maryland, 20742, USA}}
\author{Mohamad Al-Sheikhly}
\affiliation{\mbox{Nuclear Engineering Program, University of Maryland, College Park, Maryland, 20742, USA}}
\author{Alan K. Thompson}
\affiliation{\mbox{National Institute of Standards and Technology, Gaithersburg, Maryland, 20899, USA}}
\author{Robert E. Vest}
\affiliation{\mbox{National Institute of Standards and Technology, Gaithersburg, Maryland, 20899, USA}}
\author{Matthew K. Sprague}
\affiliation{\mbox{National Institute of Standards and Technology, Gaithersburg, Maryland, 20899, USA}}
\author{Karl K. Irikura}
\affiliation{\mbox{National Institute of Standards and Technology, Gaithersburg, Maryland, 20899, USA}}
\author{Charles W. Clark}
\affiliation{\mbox{National Institute of Standards and Technology, Gaithersburg, Maryland, 20899, USA}}
\affiliation{\mbox{Institute for Physical Science and Technology, University of Maryland, College Park, Maryland, 20742, USA}}
\affiliation{Joint Quantum Institute, National Institute of Standards and Technology and University of Maryland, Gaithersburg, Maryland, 20899, USA}

\date{\today}

\begin{abstract}

In the course of investigations of thermal neutron detection based on mixtures of $^{10}$BF$_3$ with other gases, knowledge was required of the photoabsorption cross sections of $^{10}$BF$_3$ for wavelengths between 135 and 205 nm.  Large discrepancies in the values reported in existing literature led to the absolute measurements reported in this communication.  The measurements were made at the SURF III synchrotron radiation facility at the National Institute of Standards and Technology.  The measured absorption cross sections vary from 10$^{-20}$ cm$^2$ at 135 nm to less than 10$^{-21}$ cm$^2$ in the region from 165 to 205 nm. Three previously unreported absorption features with resolvable structure were found in the regions 135 to 145 nm, 150 to 165 nm and 190 to 205 nm. Quantum mechanical calculations, using the TD-B3LYP/aug-cc-pVDZ variant of time-dependent density functional theory implemented in Gaussian 09, suggest that the observed absorption features arise from symmetry-changing adiabatic transitions.
\end{abstract}

\maketitle

\section{Introduction}

Boron-10 trifluouride, $^{10}$BF$_3$, was first used in a neutron detector by Amaldi \emph{et al.}, \cite{1,2} and for many years, until the availability of $^3$He, it was commonly used in gaseous proportional counters to detect thermal neutrons.  The advantage of $^{10}$BF$_3$ as a neutron detection medium is based on its relatively large thermal neutron capture cross-section, $\sigma_{\mbox{th}} = 3840$ b $(3.84 \times 10^{-21}$ cm$^2)$, for the reactions: 
\begin{eqnarray}
^{10}\mbox{B} + n & \to& ^7\mbox{Li} + \alpha + 2.792 \mbox{ MeV}, \label{eq1} \\
^{10}\mbox{B} + n & \to& ^7\mbox{Li}^* + \alpha + 2.310 \mbox{ MeV}, \label{eq2} \\
& & ^7\mbox{Li}^* \to \, ^7\mbox{Li} + \gamma \, (0.482 \mbox{ MeV}) \label{eq3}
\end{eqnarray}
Eq. (\ref{eq2}) describes the major reaction branch with a branching ratio of $94 \%$.\cite{3,4}   It is followed by prompt gamma emission according to Eq. (\ref{eq3})

In traditional BF$_3$ proportional counters, the energetic products of the reactions in Eqs. (\ref{eq1}) and (\ref{eq2}) precipitate a cascade of gas ionization in applied electric fields of approximately 100 kV/m. The charged particles are then collected as a current pulse. In contrast, we aim to build a detector similar to that described in Hughes \emph{et al.}, \cite{5} in which $^3$He was mixed with various noble gases and irradiated with cold neutrons to produce the reaction:
\begin{equation}
^3\mbox{He} + n \to ^3\mbox{H} + p + 0.765 \mbox{ MeV}.
\label{eq4}
\end{equation}
As the reaction products in Eq.\ \ref{eq4} deposit energy in the surrounding noble gas, excimers are produced. The excimers are loosely bound noble gas diatomic molecules that exist only in excited electronic states. Noble gas excimers decay by emission of vacuum ultraviolet (VUV) radiation that can be detected as a signature of neutron absorption.  Hughes \emph{et al.} \cite{5} found that tens of thousands of excimer VUV photons were generated per absorbed neutron, and in some cases approximately 30 \% of the net nuclear reaction energy was channeled into VUV emissions by this process.  McComb \emph{et al.} \cite{6} found that around ten thousand VUV excimer photons could be generated per absorbed neutron for neutrons incident upon a thin film of boron in a noble gas near atmospheric pressure. The success of these experiments was contingent upon the transparency of the heavy noble gases to their own excimer radiation. The measurements described here were motivated by concerns about VUV absorption by BF$_3$, which if sufficiently strong would disqualify BF$_3$ as a substitute for $^3$He in the arrangement described in Ref.\ \onlinecite{5}.

\section{Measurements}

Measurements of the absolute photoabsorption cross section of BF$_3$ were performed on beamline 4 (BL-4) \cite{7} of the Synchrotron Ultraviolet Radiation Facility (SURF III) at the National Institute of Standards and Technology (NIST).  The the gas handling system is shown in Fig.\ \ref{fig2} and optical system used to perform these measurements is shown in Fig.\ \ref{fig2}.

\subsection{Handling of the BF$_3$ sample}

Due to the hazards of the sample gas, several remarks are in order.  Boron has two stable isotopes, of mass number 10 and 11, with relative terrestrial abundances of approximately 19.9 \% and 80.1 \%, respectively.\cite{8}  Only $^{10}$B absorbs low energy neutrons, thus BF$_3$ gas enriched in $^{10}$B to $99.60 \pm 0.01$ \% by weight was used for these measurements. This sample had miscellaneous impurities of approximately 0.01 \%, according to the supplier's assay.\cite{9}  BF$_3$ is non-flammable; has a pungent, suffocating odor; and is a major inhalation and contact hazard with a toxicity threshold of 0.0001 \%.  It reacts with most metals with the exception of stainless steel and when exposed to moisture, borofluoric acid is formed.\cite{10}  

During the measurements, the gas handling system, shown in Fig.\ \ref{fig2}, delivered $^{10}$BF$_3$ to the photoabsorption cell.  The design of the system took full account of the toxic and corrosive properties of BF$_3$.  The absorption cell and gas handling system used materials resistant to attack by BF$_3$, and the surface area of the system was minimized to reduce BF$_3$ adsorption.  Connections between system components were made with metal seal VCR{\scriptsize \texttrademark} tube stainless steel fittings and Conflat{\scriptsize \texttrademark} flanges.\cite{23} VCR seals between fittings and flanges were stainless steel or nickel-plated copper; Conflat seals were oxygen-free, high-conductivity copper. 

Before filling the absorption cell with $^{10}$BF$_3$, it and all associated gas lines were evacuated with a liquid nitrogen trapped turbo-molecular pump while being baked.  Ultimate base pressures of $7 \times 10^{-5} \,$ Pa were routinely attained. After each set of absorption measurements, it was necessary to completely eliminate $^{10}$BF$_3$ from the absorption cell. This was done by first flushing the system with dry nitrogen to entrain the residual $^{10}$BF$_3$ and subsequently pumping the system, with the vacuum pump protected by the BF$_3$ scrubber. The entrained $^{10}$BF$_3$ then passed through a commercial BF$_3$ scrubber. , Cleanvent\textsuperscript{TM}, CO2S, BF3 from Clean Systems, Inc.\cite{23}  The scrubber was made specifically for BF$_3$ purging and has a capacity of 50 l of BF$_3$ and a certified outlet concentration of 0.0001\% BF$_3$.  A continuous Smart Gas\textsuperscript{TM} monitoring system consisting of an electrochemical sensor cell with a full-scale sensitivity of 0.0003\% BF$_3$ was located at the wall of the evacuated containment chamber closest to the $^{10}$BF3 cylinder.  The system was set to sound an alarm at BF$_3$ concentrations of 0.00005\%.  The outlet of the glove box as well as the exhausts of the pumps and the line to the absorption cell were all routed to the scrubber through a manifold.  The outlet of the scrubber, cleaned of $^{10}$BF3 was connected to the external ventilation system of the SURF III facility.

\subsection{Optical measurement procedures}

The SURF III facility delivered radiation with a bandwidth of 0.25 nm from the electron storage ring to an absorption cell at an end station on the beamline designated BL-4. \cite{11,12} A diagram of the radiation path is in Fig.\ \ref{fig2}.  Starting at the electron storage ring, radiation was imaged onto the entrance aperture of a 2-m monochromator with grazing incidence pre-optics.  Radiation leaving the monochromator through the exit aperture was imaged, by two mirrors, onto the entrance window of the absorption cell. The absorption cell was mounted on a translation stage inside an evacuated chamber.  A second chamber, located behind the exit window of the absorption cell, contained a detector for measuring the transmitted radiation.  A beam splitter in front of the first evacuated chamber reflected a small fraction of the incident light to a monitor photodiode to correct the transmitted signal for source intensity drift. BL-4 is normally operated with a larger bandpass and hence more optical power. Due to the reduced bandpass used in these experiments, the signal from the monitor photodiode was too low to use and, as discussed below, a different technique for source intensity drift correction was developed. Translation of both the absorption cell and the transmitted radiation detector enabled alignment with the incident radiation.

The absolute $^{10}$BF$_3$ photoabsorption cross section, $\sigma_\lambda$, as a function of wavelength, $\lambda$, was determined by measuring the transmission coefficient, $t_P(\lambda)$, of the gas at pressure $P$.  According to Beer's Law, 
\begin{equation}
t_P(\lambda) = \frac{I_t(\lambda)}{I_0(\lambda)} = e^{-\sigma_\lambda n l} = e^{-\sigma_\lambda (P/kT) l},
\label{eq5}
\end{equation}
where $I_t(\lambda)$ is the photodiode signal from light transmitted by the $^{10}$BF$_3$, $I_0(\lambda)$ is the photodiode signal from light transmitted by an evacuated absorption cell (assuming a constant incident optical power), $n$ is the number density of absorbing atoms or molecules, $l$ is the path length of the incident radiation through the absorption cell, $k$ is the Boltzman constant, and $T$ is the temperature.  The final expression in Eq.\ \ref{eq5} is obtained by substituting $P/kT$ for $n$ according to the ideal gas law. The pressure was measured with an industrial pressure transducer with a range of 0 to 300 kPa, an accuracy of 1.5\% for pressures $P \leq 101.3 \,$ kPa, and a linearity of 0.2\%.  The path length, $l$, of the cell was measured to better than 0.01 \%.  Comparison of the ideal gas law to the virial equation using density, pressure, and temperature reference data for BF$_3$ in the NIST Web Thermo Tables, \cite{13} showed agreement within 1 \% over the range of the experimental pressures. Because the evacuated absorption cell was used to measure $I_0(\lambda)$, knowledge of the transmissions of the MgF$_2$ entrance and exit windows of the absorption cell were not required, assuming that they remained constant over the duration of the measurements. Optical power levels used during the experiments were insufficient to darken the MgF$_2$.

Spectral scans were performed over a range of $^{10}$BF$_3$ pressures. The first scan of each series was performed with the absorption cell evacuated.  The results of this scan provided the incident signal for the transmission measurements.  Following this scan, the absorption cell was isolated from the vacuum pump, a measured pressure of $^{10}$BF$_3$ was admitted into the cell, and the spectral scan was repeated.  Subsequent scans were performed at $^{10}$BF$_3$ pressures up to 101.3 kPa. The cell was then cleared of $^{10}$BF$_3$ and evacuated to base pressure for the final scan.  This second scan of the evacuated cell was necessary for the source intensity drift correction procedure described below.

The electron beam current in the storage ring, and hence the intensity of the synchrotron radiation, decays approximately exponentially with a rate constant that can be determined from two scans of an evacuated absorption cell at two different times. The rate constant may then be used to correct measurements for  decaying radiation intensity.  The transmission of the $^{10}$BF$_3$, $t_P(\lambda)$, at a given pressure, $P$, is determined by Eq.\ \ref{eq5} using the corrected photocurrent for the input quantity $I_t(\lambda)$. The use of the correction factor introduces an uncertainty of 2.5 \% to 5 \% in the determination of $t_P(\lambda)$. Uncertainties in the pressure measurements were negligible compared to those of the correction factor.

The transmission data as a function of $^{10}$BF$_3$ pressure at each wavelength are fit to the functional form of Beer's Law to obtain values for the exponential $-\sigma_\lambda l / kT = R$. The cross section is then given by:
\begin{equation}
\sigma_\lambda = -RkT/l
\end{equation}

Seven sets of data were collected that consisted of measurements of $I_t(\lambda)$ and $I_0(\lambda)$ at intervals of 0.25 nm for a range of $^{10}$BF$_3$ pressures.  The first four sets of data, on four different dates, are shown in Fig.\ \ref{fig3}. They were collected at $^{10}$BF$_3$ pressures of 23, 45, 140, 225, 450, and 760 torr between 135 nm and 205 nm .  This region fully covers the wavelengths of the emission spectra of Ar, Kr, and Xe excimers. These data revealed previously unreported absorption features in the regions 135 nm to 145 nm, 150 nm to 165 nm, and 190 nm to 205 nm. Three additional sets of data were collected between 135 nm and 145 nm (Fig.\ \ref{fig4}) at pressures of 50, 90, 135, 160, 190, and 250 torr of $^{10}$BF$_3$, between 150 nm and 165 nm (Fig.\ \ref{fig5}) at pressures of 250, 340, 468, 570, 675, and 750 torr, and between 190 nm and 205 nm (Fig.\ \ref{fig6}) at pressures of 380, 460, 530, 600, 680, and 760 torr. 

In the analysis that follows, a resolved peak was identified by any series of at least three consecutive wavelengths where the measured cross section was greater than the background (determined from the surrounding points) by at least the standard uncertainty. Identified peaks were fitted to a Gaussian function to determine their center and width. The center of this fit was used as the wavelength for determination of the transition energy.

\section{Results}

The experimental results appear in Figs.\ \ref{fig3}-\ref{fig8}.  As shown in Fig.\ \ref{fig3}, the cross sections are small over the range of the measurements, $10^{-20}$ cm$^2$ at 130 nm falling to less than $10^{-21}$ cm$^2$ from 160 nm to 210 nm. This figure also shows the comparison of our data with the previous photoabsorption measurements of Maria, \emph{et al.}, \cite{14} Hagenow, \emph{et al.}, \cite{15} and Suto, \emph{et al.}\cite{16}

The data of Maria, \emph{et al.} cover a limited wavelength range and show some overlap with our data in the region of 155 nm to 160 nm.  Electron energy loss data by Durrant, \emph{et al.} \cite{17} show a broad feature at around 155 nm. In this region, we see a group of overlapping peaks with a maximum photoabsorption cross section around $3 \times 10^{-21}$ cm$^2$ (3000 b). The total oscillator strength associated with the 155 nm to 160 nm region is $\approx 10^{-5}$,  characteristic of a weak transition.\cite{17} Maria, \emph{et al.} \cite{14} characterize their absorption spectrum as evincing the excitation of a triplet state from the singlet ground state, which would be weak in molecules consisting of first-row elements.  They find analogous features in the absorption spectra of the heavier boron trihalides BCl$_3$, BBr$_3$ and BI$_3$, which grow considerably in strength with increasing atomic number, as would be expected for triplet excitation.

The Hagenow, \emph{et al.} \cite{15} data come from a study that focuses on photoabsorption at shorter wavelengths, and the longest-wavelength distinctive feature they report lies at 95 nm, which is outside the reach of our instrument.  That feature has a peak cross section of $4.5 \times 10^{-17}$ cm$^2$ (45 Mb) and an integrated oscillator strength of $\sim 0.26$, characteristic of a strong allowed transition.  Their data at longer wavelengths shows no distinctive features, and it is not clear that their experiment had the dynamic range required to resolve features with cross sections of 0.0003 Mb such as we see here.

Our summation of the Hagenow, \emph{et al.} \cite{15} data comes from digitization of the line drawing in the first figure of their paper. Those data are not accompanied by any statement of uncertainty from the original authors.

The studies of Suto, \emph{et al.} \cite{16} cover the wavelength range 45 nm to 113 nm, which again is outside the range of our instrument.  However, they estimate the cross section at wavelengths longer than 120 nm to be less than $10^{-21}$ cm$^2$.  In a study of vacuum-ultraviolet absorption spectra of the boron trihalides, Planckaert, \emph{et al.} \cite{19} state that ``BF$_3$ does not have any absorption bands up to 85000 cm$^{-1}$,'' i.e. at wavelengths longer than 118 nm.

A large scale plot of the photoabsorption data taken on four different days is shown in Fig.\ \ref{fig4}. Of special note are the three regions between 135 nm and 145 nm, 150 nm and 165 nm, and 190 nm to 204 nm. These regions, shown in Figs.\ \ref{fig5}, \ref{fig6}, and \ref{fig7}, and summarized in Table \ref{Levels} and Fig.\ \ref{fig8}, show structure in the photoabsorption spectrum. None of the structure is related to the major impurities identified by the manufacturer: CO$_2$ and SiF$_4$, both present in concentrations of 0.0058 \% (see Appendix).  Neither of these structures are related to O$_2$, N$_2$, or H$_2$O.  

\begin{table}[h]
\caption{Energy levels of $^{10}$BF$_3$ as inferred from photoabsorption data in the wavelength range 135- 205 nm.  We assign a uniform uncertainty to the wavelength values of $\pm 0.25\,$ nm, derived from the resolution of the spectrometer. The energies of the inferred levels with respect to the $^{10}$BF$_3$ ground state are given in eV, and are depicted in Fig. \ref{fig8} }
\label{Levels}
\begin{tabular}{l c}
\hline \hline

Wavelength (nm) & Energy (eV) \\ \hline

203.37	&	6.10	\\
201.70	&	6.15	\\
200.16	&	6.19	\\
198.62	&	6.24	\\
197.04	&	6.29	\\
195.59	&	6.34	\\
193.97	&	6.39	\\
159.95	&	7.75	\\
157.37	&	7.88	\\
154.56	&	8.02	\\
140.76	&	8.81	\\
139.41	&	8.89	\\
137.88	&	8.99	\\

\hline \hline
\end{tabular}
\end{table}

\section{Analysis}

To identify the origins of the observed absorption features in the regions around 140 nm (8.9 eV), 160 nm (7.7 eV), and 200 nm (6.2 eV), calculations of the spectral line positions for both vertical and adiabatic electronic excitation of BF$_3$ were undertaken.  Additionally, vibrational excitation energies were calculated for the ground state and several of the excited states with C$_s$ symmetry. 

The calculations used the TD-B3LYP/aug-cc-pVDZ variant of time-dependent density functional theory,\cite{20} as implemented in Gaussian 09.\cite{21} This model chemistry is expected to yield vertical and adiabatic excitation energies to within $\pm 1 $ eV (95\% coverage factor).\cite{22} Here we provide just a summary of results, details of which will be provided in a later report.

\begin{table*}[t]
\caption{Vertical excitation energies of states of $^{10}$BF$_3$, calculated at TD-B3LYP/aug-cc-pVDZ.}
\label{Vertical}
\begin{tabular}{c c c c | c c c c}
\hline \hline

\multicolumn{4}{c|}{Singlet States} & \multicolumn{4}{c}{Triplet States} \\ \hline

Electronic State & Symmetry & $E$ (eV) & $\lambda$ (nm) & Electronic State & Symmetry & $E$ (eV) & $\lambda$ (nm) \\ \hline

$^1$E$'$ & D$_{\mbox{3h}}$ & 11.34 & 109 & $^3$E$'$ & D$_{\mbox{3h}}$ &11.12 & 111 \\
$^1$E$''$ & D$_{\mbox{3h}}$ & 11.29 & 110 & $^3$E$''$ & D$_{\mbox{3h}}$ & 11.04 & 112 \\
$^1$E$''$ & D$_{\mbox{3h}}$ & 11.07 & 112 & $^3$E$''$ & D$_{\mbox{3h}}$ & 10.85 & 114 \\
$^1$A$_2'$ & D$_{\mbox{3h}}$ & 10.27 & 121 & $^3$E$'$ & D$_{\mbox{3h}}$ & 10.47 & 118 \\
$^1$A$_1''$ & D$_{\mbox{3h}}$ & 10.24 & 121 & $^3$A$_2'$ & D$_{\mbox{3h}}$ & 10.08 & 123 \\
$^1$A$_1'$ & D$_{\mbox{3h}}$ & 0.00 & Ground & $^3$A$_1''$ & D$_{\mbox{3h}}$ & 9.99 & 124 \\
\hline \hline
\end{tabular}
\end{table*}

\begin{table*}[t]
\caption{Adiabatic excitation energies of states of  $^{10}$BF$_3$, calculated at TD-B3LYP/aug-cc-pVDZ.}
\label{Adiabatic}
\begin{tabular}{c c c c | c c c c}
\hline \hline

\multicolumn{4}{c|}{Singlet States} & \multicolumn{4}{c}{Triplet States} \\ \hline

Electronic State & Symmetry & $E$ (eV) & $\lambda$ (nm) & Electronic State & Symmetry & $E$ (eV) & $\lambda$ (nm) \\ \hline

$^1$A$'$ & C$_{\mbox{s}}$ & 10.95 & 113 & $^3$A$'$ & C$_{\mbox{s}}$ & 10.16 & 122 \\
$^1$A$'$ & C$_{\mbox{s}}$ & 10.52 & 118 & $^3$A$_1$ & C$_{\mbox{3v}}$ & 8.87 & 140 \\
$^1$A$'$ & C$_{\mbox{s}}$ & 8.79 & 141 & $^3$A$_2$ & C$_{\mbox{2v}}$ & 8.53 & 145 \\
$^1$A$_2$ & C$_{\mbox{2v}}$ & 8.56 & 145 &  &  &  &  \\
$^1$A$'$ & C$_{\mbox{s}}$ & 7.21 & 172 &  &  &  &  \\
$^1$A$''$ & C$_{\mbox{s}}$ & 7.02 & 177 & $^3$A$''$ & C$_{\mbox{s}}$ & 6.74 & 184 \\
$^1$A$'$ & C$_{\mbox{s}}$ & 6.99 & 177 & $^3$A$''$ & C$_{\mbox{s}}$ & 6.56 & 189 \\
$^1$A$_1'$ & D$_{\mbox{3h}}$ & 0.00 & Ground &  &  &  &  \\
\hline \hline
\end{tabular}
\end{table*}

\begin{table*}[t]
\caption{Adiabatic excitation energies of states of  $^{10}$BF$_3$ shifted by 0.7 eV and their associated harmonic vibrational energies, calculated at TD-B3LYP/aug-cc-pVDZ.}
\label{AdEx}
\begin{tabular}{c c c c c c c c c c}
\hline \hline


\multirow{2}{*}{State} & \multirow{2}{*}{Symmetry} & \multirow{2}{*}{$E$ (eV)} & \multirow{2}{*}{$\lambda$ (nm)} & \multicolumn{6}{c}{Vibrational Energies (eV) and Symmetries} \\ \cline{5-10}

& & & & 1 & 2 & 3 & 4 & 5 & 6 \\ \hline


$^1$A$'$ & C$_{\mbox{s}}$ & 7.21 & 172 & 0.036(A$'$) & 0.109(A$'$) & 0.110(A$'$) & 0.121(A$''$) & 0.231(A$'$) & 0.372(A$'$) \\
$^1$A$''$ & C$_{\mbox{s}}$ & 7.02 & 177 & 0.026(A$'$) & 0.029(A$'$) & 0.031(A$'$) & 0.077(A$'$) & 0.134(A$'$) & 0.191(A$''$) \\
$^1$A$'$ & C$_{\mbox{s}}$ & 6.99 & 177 & 0.018(A$''$) & 0.044(A$''$) & 0.051(A$''$) & 0.067(A$'$) & 0.114(A$'$) & 0.153(A$'$) \\
$^3$A$''$ & C$_{\mbox{s}}$ & 6.74 & 184 & 0.021(A$''$) & 0.040(A$''$) & 0.051(A$'$) & 0.066(A$'$) & 0.112(A$'$) & 0.151(A$'$) \\
$^3$A$''$ & C$_{\mbox{s}}$ & 6.56 & 189 & 0.018(A$''$) & 0.019(A$'$) & 0.035(A$'$) & 0.065(A$'$) & 0.132(A$'$) & 0.179(A$''$) \\
$^1$A$_1'$ & D$_{\mbox{3h}}$ & 0 & Ground & 0.056(E$'$) & 0.056(E$''$) & 0.087(A$_2''$) & 0.106(A$_1'$) & 0.175(E$'$) & 0.175(E$'$) \\

\hline \hline
\end{tabular}
\end{table*}

Table \ref{Vertical} lists the calculated vertical excitations to both singlet and triplet excited states from the $^1$A$_1'$ ground state.  The ground state of BF$_3$ is of D$_{3h}$ symmetry as are the symmetries of the excited states listed there.  The longest-wavelength transition from the $^1$A$_1'$ ground state to the lowest $^3$A$_1''$ triplet state is 124 nm (9.99 eV), much shorter than the observed absorption wavelengths. We conclude that the observed absorption spectra are unlikely to arise from vertical transitions. Table \ref{Adiabatic} lists the calculated adiabatic electronic energy levels, point groups, and electronic symmetries for several electronically-excited states of BF$_3$.  Here, the lowest energy transitions are at 189 nm (6.56 eV) for the lowest energy triplet and 177 nm (6.99 eV) for the lowest energy singlet. The observed features around 200 nm (6.2 eV) have the character of a vibration series in the upper state of a symmetry-changing adiabatic transition. Table \ref{AdEx} shows the calculated vibration level spacing for the ground state and for several of the C$_s$ symmetry excited states.

The assignment of calculated transitions to the observed absorption features is imperfect. The calculations do not correspond one-to-one with the observations. Given the approximately 1 eV uncertainty of the calculations, the need to shift the calculated energies is reasonable.\cite{22} However, no global shift in transition energies is adequate to match the calculated and observed features. Nonetheless, it is important to note that there is good theoretical reason to expect absorption features in the spectrum of BF$_3$ in this wavelength region as a result of adiabatic transitions.

\section{Conclusions}

The $^{10}$BF$_3$ absolute photoabsorption cross section has been measured  at the SURF III facility with an overall uncertainly of 20 \% for wavelengths between 135 and 205 nm. The major contributor to the uncertainty is the decay of the synchrotron ring current over the course of the measurement period.   The measured cross sections range from 10$^{-20}$ cm$^2$ at 135 nm to 10$^{-21}$ cm$^2$ over the region from 165 to 205 nm. Three regions with resolvable structure have been observed for the first time around 140 nm (8.9 eV), 160 nm (7.7 eV), and 200 nm (6.2 eV).

In an attempt to identify these features, quantum mechanical calculations using the TD-B3LYP/aug-cc-pVDZ variant of time-dependent density functional theory, implemented in Gaussian 09 have been performed. Calculations to excited states with the same D$_{3h}$ symmetry as the ground state show no correspondence to the experimental results. Adiabatic transitions to different symmetry states of the BF$_3$ molecule, however, may account for the  structure. Assignment of the observed features to specific calculated transitions was not entirely successful.

\section*{Acknowledgements}

We benefitted from conversations with Michael J. Brunger and Jon Hougen, and thank the U.S. Nuclear Regulatory Commission (NRC) for its financial support (NRC fellowship grant NRC-38-09-934).

\appendix*

\section{Gas sample analysis}

Trace contamination is a potential cause of the three structures in the reported photoabsorption data between 135 nm and 145 nm, 150 nm and 165 nm, and 190 nm to 205 nm. The $^{10}$BF$_3$ specified to have a chemical purity of greater than 99.9 \% and an isotopic enrichment of greater than 99 \%.  Analysis of the gas by the supplier using plasma mass spectrometry, gas chromatography, and Fourier transform infrared spectroscopy returned the impurity budget shown in Table \ref{Impurities}. None of the listed impurities correspond to the observed structures. We conclude that the observed structures are not likely to originate from impurities.

\begin{table}[h]
\caption{Impurity budget.}
\label{Impurities}
\begin{tabular}{l c}
\hline \hline

Impurity & Impurity Level (ppm) \\ \hline
Nitrogen & 4.62 $\pm$ 1 \\
Oxygen + Argon & 1.33 $\pm$ 1 \\
Carbon Dioxide & 57.57 $\pm$ 2 \\
Sulfur Dioxide & 8.60 $\pm$ 1 \\
Silicon Tetrafluoride & 57.56 $\pm$ 1 \\
Hydrogen Fluoride & 1.00 $\pm$ 1 \\

\hline \hline
\end{tabular}
\end{table}

\begin{figure*}
\includegraphics[width=15cm,height=15cm]{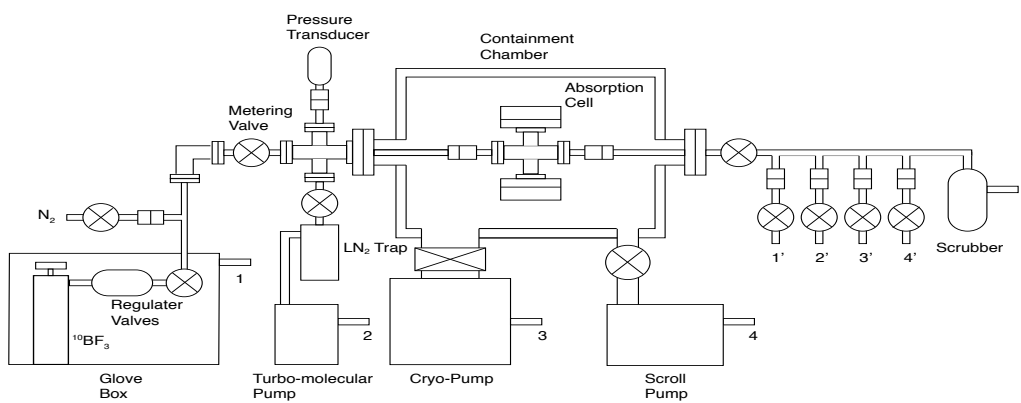}
\caption{BF$_3$ gas handling system and absorption cell.  The $^{10}$BF$_3$ gas cylinder
was enclosed in a sealed glove box.  The absorption cell was enclosed in a containment chamber that was evacuated by scroll and cryogenic pumps. All lines exposed to the gas were evacuated with a liquid nitrogen trapped turbo-molecular pump. $^{10}$BF$_3$ was introduced into the absorption cell through a metering valve connected to the corrosive gas regulator valve on the $^{10}$BF$_3$ cylinder.  An industrial pressure transducer measured the pressure in the absorption cell through which the EUV radiation passed. $^{10}$BF$_3$ was removed from the system by first flushing with dry nitrogen and then pumping with the turbo molecular pump. The exhausts from the pumps (2, 3, and 4), the glove box (1), and the nitrogen flushing lines were connected to a BF$_3$ scrubber via lines 1$'$, 2$'$, 3$'$, and 4$'$ before being directed to the high velocity ventilation duct of the facility.}
\label{fig1}
\end{figure*}

\begin{figure*}
\centering
\includegraphics[width=\textwidth]{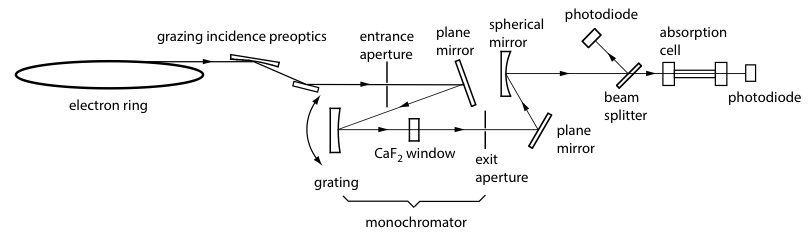}
\caption{Schematic diagram of BL-4 at the NIST SURF III facility.  Broadband synchrotron radiation is directed by grazing incidence preoptics through the entrance aperture of a grating monochromator and on to the grating after reflection from a plane mirror.  A CaF$_2$ window separates the UHV storage ring from the down-stream experiment.  Diffracted radiation exiting the window then passes through a variable width exit aperture and is directed to the absorption cell of the experiment via plane and spherical mirrors.  A beam splitter at the entrance to the experiment sends a small fraction of the incident radiation to a photodiode that monitors the incident radiation.  A similar photodiode at the exit of the absorption cell measures the transmitted radiation.}
\label{fig2}
\end{figure*}

\begin{figure}
\includegraphics[width=\textwidth]{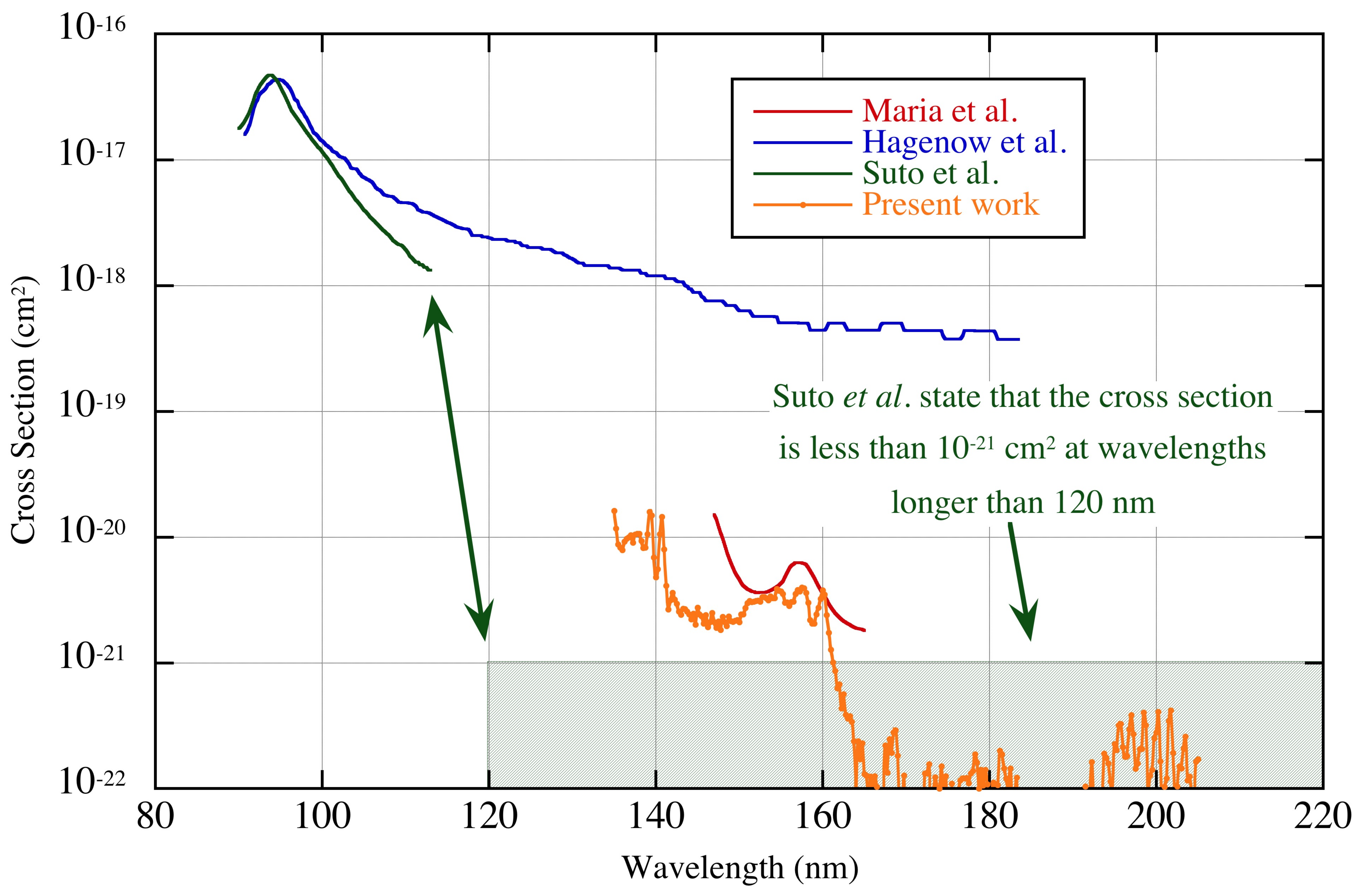}
\caption{Composite $^{10}$BF$_3$ absorption spectrum showing present results (yellow) along with previous results of Maria, \emph{et al.}\cite{14}, Hagenow, \emph{et al.}\cite{15}, and Suto, \emph{et al.}\cite{16}}
\label{fig3}
\end{figure}

\begin{figure}[t]
\centering
\includegraphics[width=\textwidth]{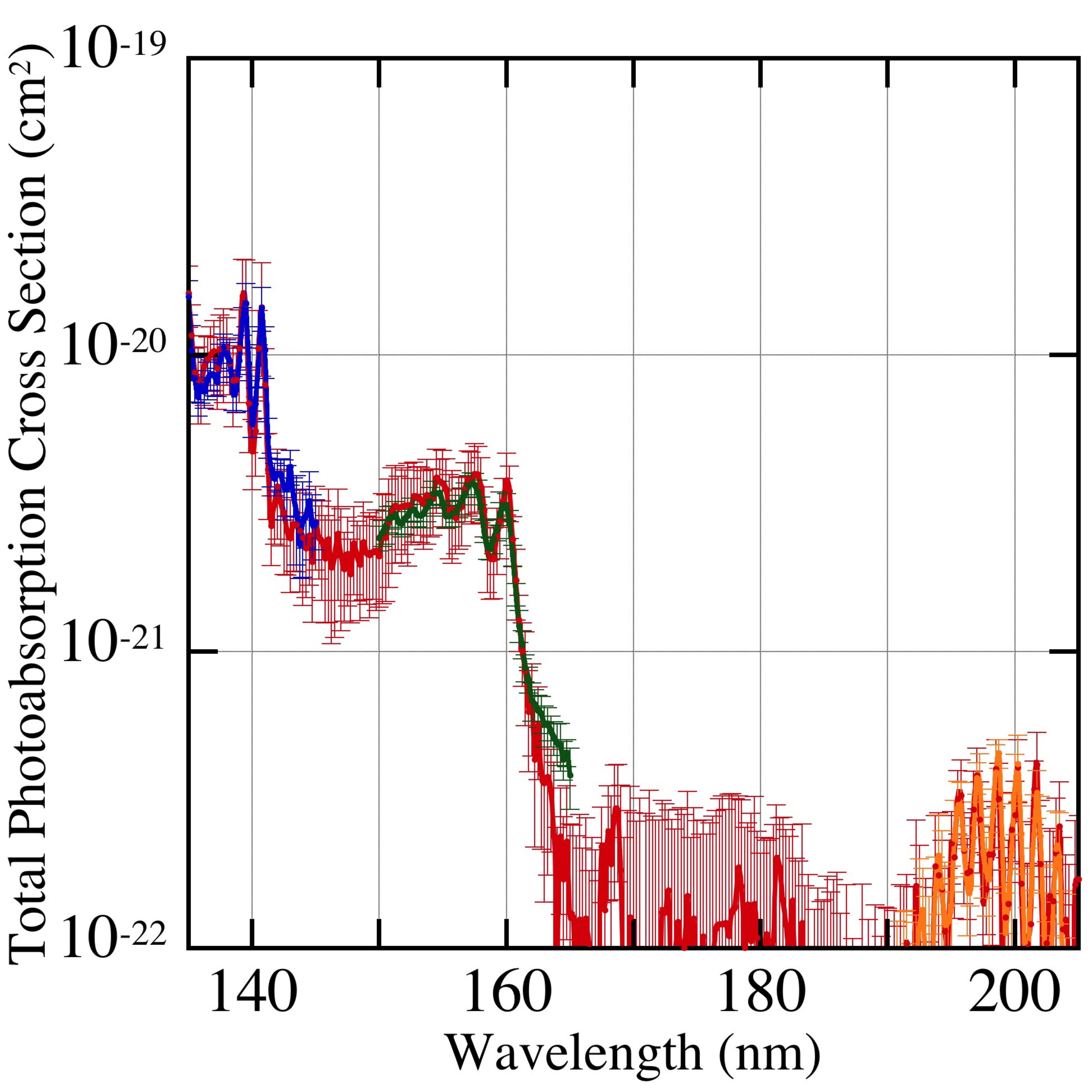}
\caption{Absolute photoabsorption cross sections for $^{10}$BF$_3$ taken at four separate dates on BL-4 at the NIST SURF III facility.}
\label{fig4}
\end{figure}
\begin{figure}[t]
\centering
\includegraphics[width=\textwidth]{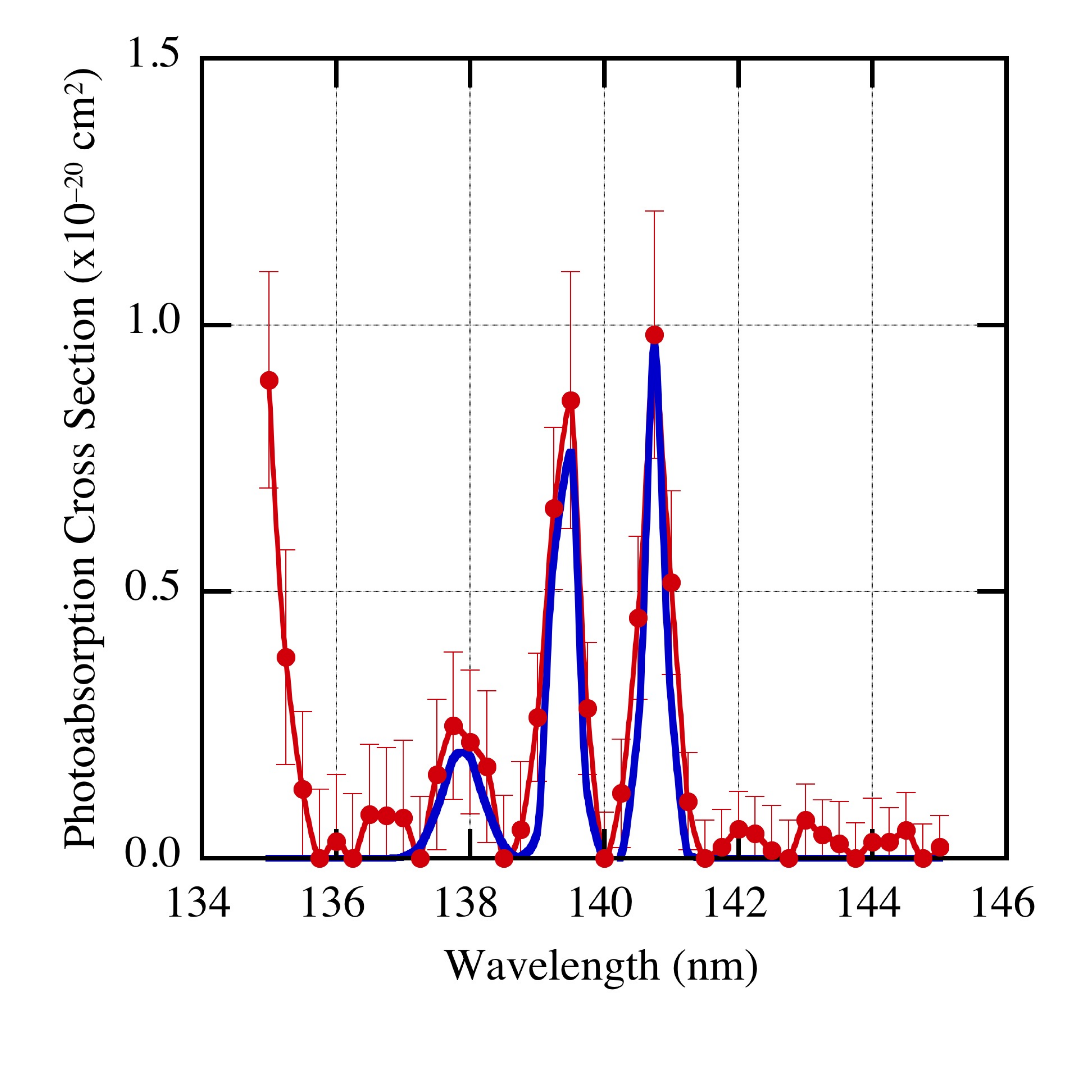}
\caption{$^{10}$BF$_3$ absorption spectrum from 135 to 145 nm showing experimental data with uncertainties and Gaussian fits to the data.}
\label{fig5}
\end{figure}

\begin{figure}[t]
\centering
\includegraphics[width=\textwidth]{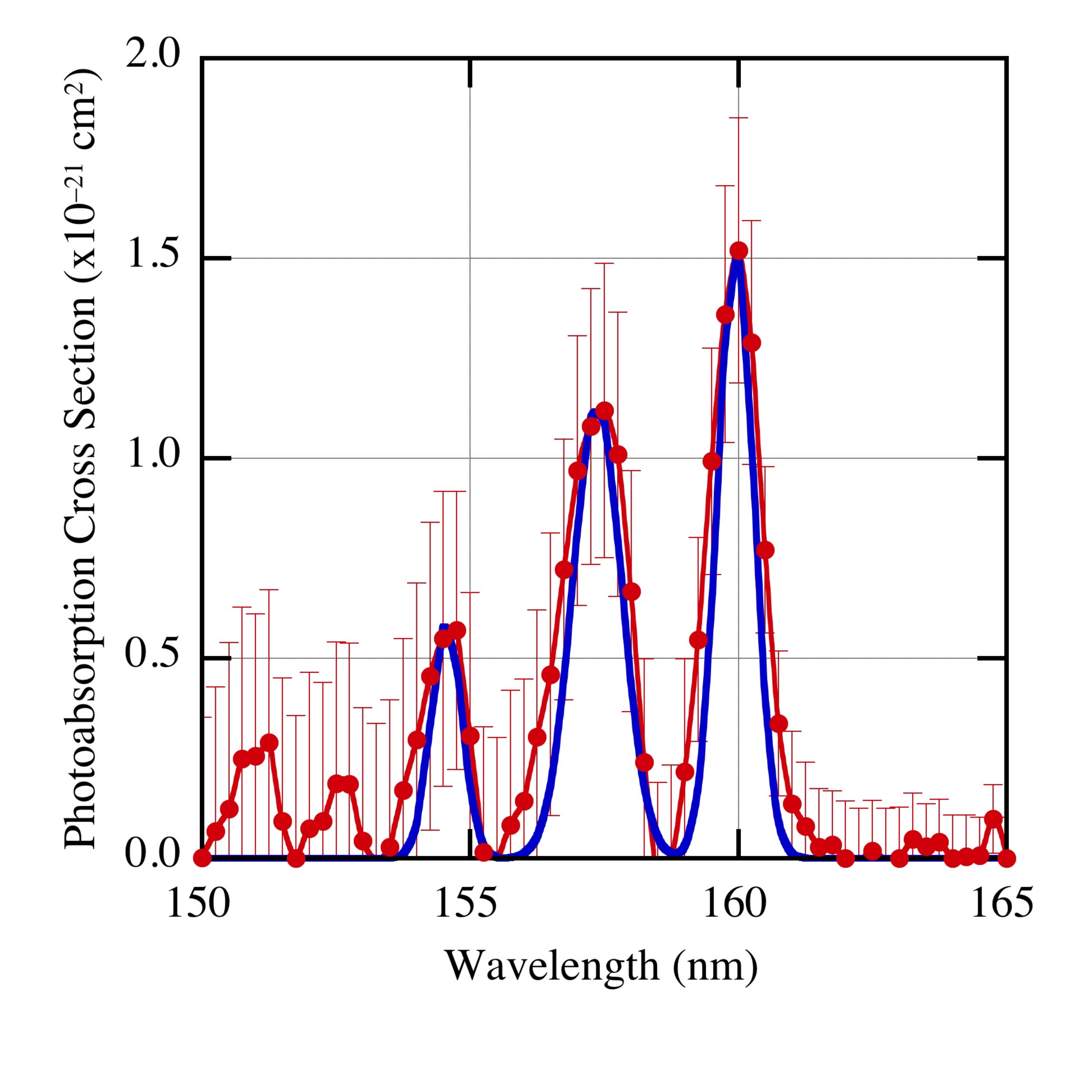}
\caption{$^{10}$BF$_3$ absorption spectrum from 150 to 165 nm showing experimental data with uncertainties and Gaussian fits to the data.}
\label{fig6}
\end{figure}
\begin{figure}[t]
\centering
\includegraphics[width=\textwidth]{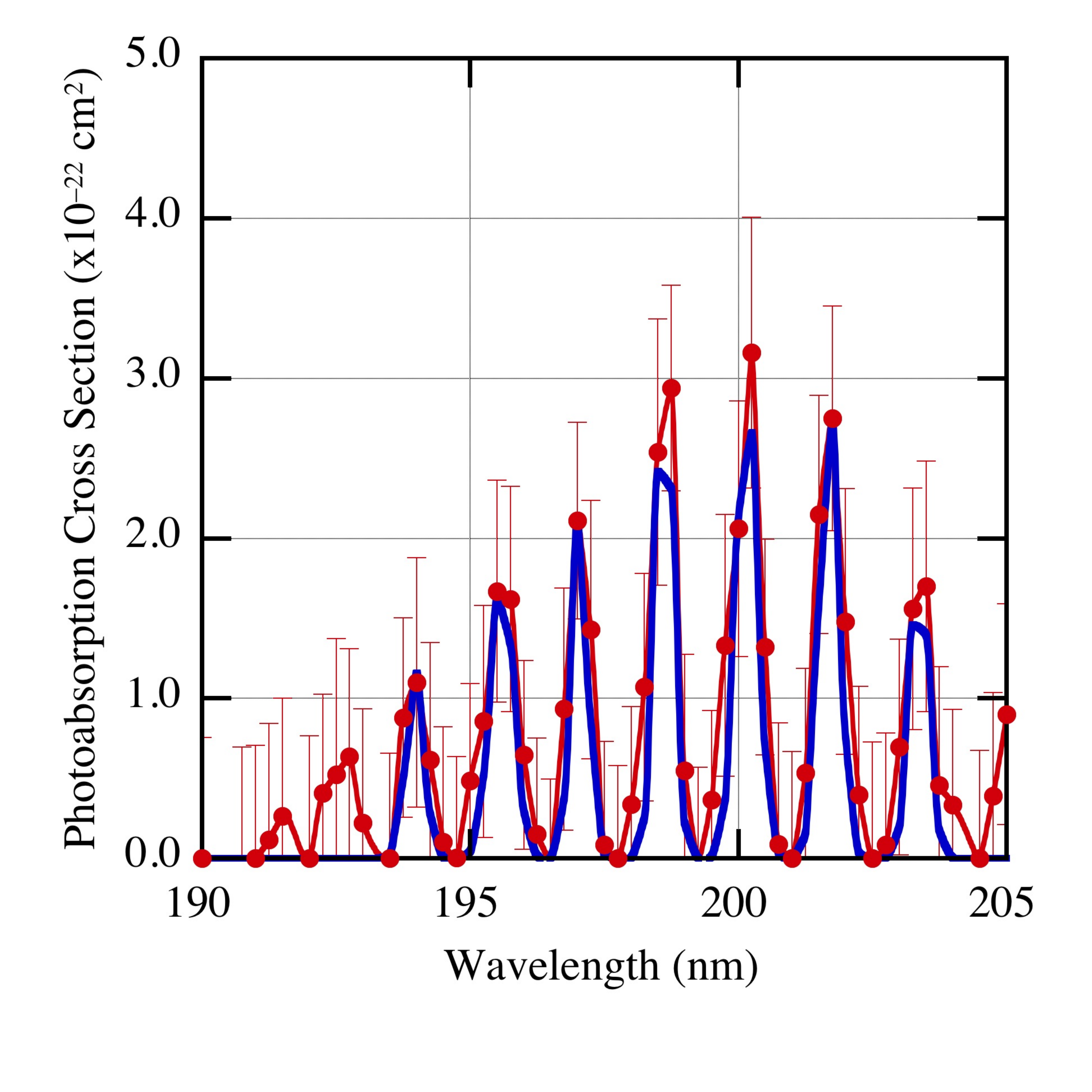}
\caption{$^{10}$BF$_3$ absorption spectrum from 190 to 205 nm showing experimental data with uncertainties and Gaussian fits to the data.}
\label{fig7}
\end{figure}

\begin{figure}[t]
\centering
\includegraphics[width=\textwidth]{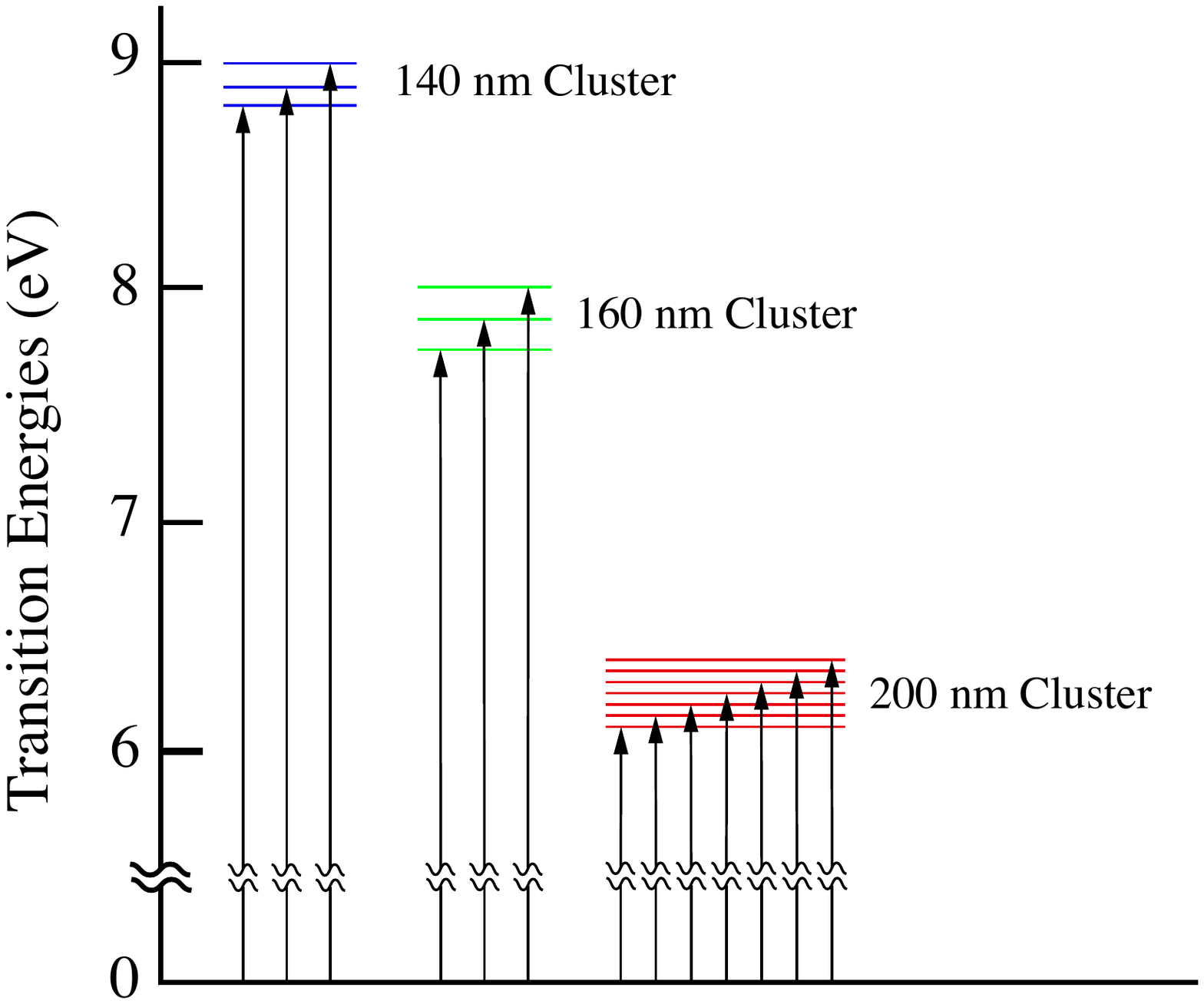}
\caption{Energy level diagram based on the analysis of the spectra in Figs.\ \ref{fig4}, \ref{fig5}, and \ref{fig6}.}
\label{fig8}
\end{figure}


\begin{thebibliography}{22}
\bibitem{23}
Certain commercial equipment, instruments, or materials are identified in this paper to foster understanding. Such identification does not imply recommendation or endorsement by the National Institute of Standards and Technology, nor does it imply that the materials or equipment identified are necessarily the best available for the purpose.
\end{thebibliography}
\end{document}